\documentclass[%
reprint,
amsmath,amssymb,prl,
aps, groupedaddress,superscriptaddress
]{revtex4-1}

\usepackage{graphicx}
\usepackage{dcolumn}
\usepackage{bm}
\usepackage{braket}
\usepackage{mathtools}
\usepackage{bbm}
\usepackage{comment}
\usepackage{makecell}
\usepackage{afterpage}

\newcommand\blankpage{%
    \null
    \thispagestyle{empty}%
    \addtocounter{page}{-1}%
    \newpage}

\usepackage{hyperref}
\hypersetup{
	colorlinks=true,
	linkcolor=blue,
	citecolor=blue,
	filecolor=magenta,
	urlcolor=cyan
}

\graphicspath{{../}}

\begin{document}

\def\papertitle{{Symmetry breaking and error correction in open quantum systems}}
\def\authornames{Simon Lieu,$^{1,2}$ Ron Belyansky,$^{1,2}$ Jeremy T.~Young,$^{1}$ Rex Lundgren,$^{1,2}$ Victor V.~Albert,$^3$ Alexey V.~Gorshkov$^{1,2}$}
\def\umd{Joint Quantum Institute, NIST/University of Maryland, College Park, MD 20742, USA}
\def\qcs{Joint Center for Quantum Information and Computer Science, NIST/University of Maryland, College Park, Maryland 20742 USA}
\def\ct{Institute for Quantum Information and Matter and Walter Burke Institute for Theoretical Physics,
California Institute of Technology, Pasadena, CA 91125, USA}

\title{\papertitle}
\author{Simon Lieu}
\affiliation{\umd}
\affiliation{\qcs}
\author{Ron Belyansky}
\affiliation{\umd}
\affiliation{\qcs}
\author{Jeremy T.~Young}
\affiliation{\umd}
\author{Rex Lundgren}
\affiliation{\umd}
\affiliation{\qcs}
\author{Victor V.~Albert}
\affiliation{\ct}
\author{Alexey V.~Gorshkov}
\affiliation{\umd}
\affiliation{\qcs}

\date{\today}

\begin{abstract}

Symmetry-breaking transitions  are a well-understood phenomenon of closed quantum systems in quantum optics, condensed matter, and high energy physics.  However, symmetry breaking in open systems is less thoroughly understood, in part due to the richer steady-state and symmetry structure that such systems possess. For the prototypical open system---a Lindbladian---a unitary symmetry can be imposed in a ``weak'' or a ``strong'' way. We characterize the possible  $\mathbb{Z}_n$ symmetry breaking transitions for both cases. In the case of $\mathbb{Z}_2$, a weak-symmetry-broken phase guarantees at most a classical bit steady-state structure, while a strong-symmetry-broken phase admits a partially-protected steady-state qubit. Viewing photonic cat qubits through the lens of strong-symmetry breaking, we show how to dynamically recover the logical information after any gap-preserving strong-symmetric error; such recovery becomes perfect exponentially quickly in the number of photons. Our study forges  a connection  between driven-dissipative phase transitions and error correction.

\end{abstract}

\maketitle

While an open quantum system typically evolves toward a  thermal state \cite{breuer2002}, non-thermal steady states emerge in the presence of an external drive \cite{noh2016, diehl2008} or via reservoir engineering \cite{zoller1996, plenio2002}. In particular, systems with multiple steady states  have recently attracted much attention due to their ability to remember initial conditions \cite{prosen2012, albert2014, albert2016, buca2019, roberts2020, mac2016, eze2019,  dutta2020, hafezi2019, moos2018, gau2020, zhang2020}.  For Markovian environments,  this involves studying Lindblad superoperators (Lindbladians)  \cite{lindblad1976,Gorini1976a, popov1969} that possess multiple eigenvalues of zero \cite{thesis}.

On the one hand, Lindbladians with such degenerate steady states are the key ingredient for  \textit{passive} error correction \cite{lidar1998, terhal2015,  mirrahimi2014, puri2017, kapit2016, passive1, passive2, passive3, passive4, passive5, passive6, passive7}. In this paradigm, the degenerate steady-state structure of an appropriately engineered Lindbladian stores the logical information, and the Lindbladian passively protects this information from certain errors by continuously mapping any leaked information back into the structure without distortion. An important task remains to identify \textit{generic} systems that host such protected qubit steady-state structures, and classify the errors that can be corrected in this way.

On the other hand, the presence of a ground-state degeneracy in the infinite-size limit of a closed system is  a salient feature of symmetry breaking (e.g.~the ferromagnetic  ground states of the Ising model) \cite{sachdev2007}. While the study of analogous phase transitions in open systems has become a rich and active field \cite{diehl2008,kirton2019, theory-dis1, theory-dis2, theory-dis3, theory-dis4, theory-dis5,  maghrebi2016,young2020,lundgren2019, keeling2013, rossini2018, exp1, rota2019, wouters2020} with significant experimental relevance \cite{exp2, exp3, exp4, exp-diss2, exp-diss2, esslinger2013}, attention has focused on the steady-state degeneracy in  symmetry-broken phases only recently \cite{ciuti2018, eisert2017, cirac2012}. 

Since steady-state degeneracy is a requirement for both passive error correction and symmetry breaking, it is natural to ask whether there are any connections between the two phenomena.  Here, we begin to shed light on this interesting and important direction by (A) describing how the dimension and structure of the steady-state manifold changes across a dissipative phase transition, and (B) identifying any passive protection due to the symmetry-broken phase (we will often drop the word symmetry  below). 

To this end, we emphasize an important distinction between ``weak'' and ``strong''  transitions which is unique to open systems. This difference stems from the dissipative part of the Lindbladian which can respect a symmetry in two  separate ways, as first noted by Bu\v{c}a and Prosen  \cite{prosen2012}. We show that the $\mathbb{Z}_2$ strong-broken phase encodes a qubit in its steady-state structure in the infinite-size limit, and that errors  preserving this structure can be passively corrected. Our analysis is made concrete by considering a driven-dissipative photonic  mode---a minimal model for the study of both non-equilibrium transitions \cite{ciuti2018} and bosonic error-correcting codes \cite{mirrahimi2014}.

\textit{Generic $\mathbb{Z}_n$ symmetry breaking.---}We consider open systems governed by  a Lindblad master equation
\begin{equation}
    \frac{d\rho}{dt} =  \mathcal{L}(\rho) = -i [H, \rho] +  \sum_i \left( 2 L_i \rho L_i^\dagger - \{L^\dagger_i L_i, \rho \}\right),
\end{equation}
with density matrix  $\rho$, Hamiltonian $H$, dissipators $L_{i}$, and Lindbladian $\mathcal{L}$. A \textit{strong symmetry} is satisfied if there exists an operator $P$ such that $[H,P]=[L_{i},P]=0,  \forall i $. A \textit{weak symmetry} is satisfied if $\left[\mathcal{L},\mathcal{P} \right]=0$, where $\mathcal{P}(\cdot) = P (\cdot) P^\dagger$. We will showcase differences between previously studied weak-symmetry transitions and the strong-symmetry ones we introduce here, focusing on changes to the dimension \textit{and} structure of the steady-state manifold.

\begin{table}
\begin{center}
\begin{tabular}{|c|c|c|c|}
\hline 
$\mathbb{Z}_n$ sym. & definition& sufficient condition & s.s.~transition \tabularnewline
\Xhline{3\arrayrulewidth}
strong & $\left[\mathcal{L},\mathcal{P}_{l,r}\right]=0$ & $[H,P]=[L_i,P]=0$& $n$-to-$n^2$  \tabularnewline
\hline 
weak  & $[\mathcal{L},\mathcal{P}]=0$  & $[H,P]=\{L_i,P\}=0$ & $1$-to-$n$ \tabularnewline
\hline 
\end{tabular}
\end{center}
\caption{Comparison of  a strong vs.~weak $\mathbb{Z}_n$ symmetry of $\mathcal{L}$. The final column describes transitions in the dimension of the steady state (s.s.) manifold (number of zero eigenvalues of $\mathcal{L}$) when going from the unbroken phase to the broken phase.} 
\label{tab:sym}
\end{table}

Let us review \cite{ciuti2018} weak $\mathbb{Z}_2$-symmetry breaking, which is similar to conventional closed-system symmetry breaking and is ubiquitous in open systems \cite{keeling2013, rossini2018,kirton2019}. Here, $P$ is a parity operator that satisfies $P|\pm\rangle=\pm|\pm\rangle$ with parity eigenvalues $\pm1$ and sets of eigenstates $\{|\pm\rangle\}$. Its superoperator version, $\mathcal{P}(\cdot)=P(\cdot)P^\dagger$, possesses $+ 1$ and $- 1$ ``superparity'' eigenvalues, belonging respectively to eigenoperators $|\pm\rangle\langle\pm|$ and $|\pm\rangle\langle\mp|$. A weak $\mathbb{Z}_2$ symmetry $\cal{P}$ can thus be used to block-diagonalize $\mathcal{L}$ into two sectors, $\mathcal{L}=\text{Diag}[\mathcal{L}_{+},\mathcal{L}_{-}]$, one for each superparity. Since the $-1$ superparity sector contains only traceless eigenoperators, the (trace-one) steady state of a finite-size system will necessarily have superparity $+1$ and be an eigenoperator of $\cal{L}_+$. If a symmetry-broken order parameter is to acquire a non-zero steady-state expectation value in the infinite-size limit, $\cal{L}_-$ must also pick up a zero-eigenvalue eigenoperator, and positive/negative mixtures of the original steady state and this new eigenoperator will become the two steady states of the system (a ``1-to-2'' transition).

In the strong case, there are two superparity superoperators, $\mathcal{P}_{l}(\cdot)=P(\cdot)$ and $\mathcal{P}_{r}(\cdot)=(\cdot)P^\dagger$, that commute with each other as well as with $\cal{L}$. Their eigenvalues further resolve the states $|+\rangle\langle+|$ from $|-\rangle\langle-|$ (and similarly $|+\rangle\langle-|$ from $|-\rangle\langle+|$), yielding the finer block diagonalization $\mathcal{L}=\text{Diag}[\mathcal{L}_{++},\mathcal{L}_{--},\mathcal{L}_{+-},\mathcal{L}_{-+}]$. The key observation is that both $\cal{L}_{++}$ and $\cal{L}_{--}$ have to admit steady-state eigenoperators, since their respective sectors house eigenoperators with nonzero trace. 
A strong  transition is therefore a 2-to-4 transition: the dimension of the steady-state manifold increases from 2 to 4 as $\mathcal{L}_{-+}$ and $\mathcal{L}_{+-}$ pick up zero eigenvalues in the broken phase. This reasoning generalizes to $\mathbb{Z}_n$ symmetries (see Table~\ref{tab:sym}).

\textit{Steady-state structure in different $\mathbb{Z}_2$ phases.---}Apart from differences in the dimension of the steady-state manifold,  a weak-broken $\mathbb{Z}_2$ phase can yield at most a classical bit structure, while a strong-broken phase can yield a qubit steady-state manifold. To see this, we express the steady state of a $\mathbb{Z}_2$-symmetric model in the parity basis,  $| \vec{\pm} \rangle = (| \pm \rangle_1,  |\pm \rangle_2,\ldots )$, as
\begin{equation} \label{eq:parity}
\rho_{ss} = 
\left(\begin{array}{cc}
s_{++} & s_{+-}\\
s_{-+} & s_{--}\\
\end{array}\right).
\end{equation}
Table \ref{tab:dof} lists the ``degrees of freedom'' for the steady state in each phase, i.e.~which part of the matrix is allowed to change depending on the initial condition $\rho_i$. The strong-broken phase can remember both the relative magnitude and phase of an initial state, which guarantees that a qubit can be encoded into the steady state. The strong-unbroken and weak-broken phases both host a classical bit structure, where classical mixtures remain stable. The weak-unbroken phase will generically possess a unique steady state.

\begin{table}
\begin{center}
\begin{tabular}{|c|c|c|}
\hline 
 $\mathbb{Z}_2$ phase & s.s.~freedom & s.s.~structure \tabularnewline
\Xhline{3\arrayrulewidth}
strong, broken & $s_{++}, s_{--}, s_{+-}, s_{-+}$  & qubit \tabularnewline
\hline 
strong, unbroken & $s_{++}, s_{--}$ & classical bit \tabularnewline
\hline 
weak, broken & $ s_{+-},s_{-+}$& classical bit \tabularnewline
\hline 
weak, unbroken & none& unique  \tabularnewline
\hline 
\end{tabular}
\end{center}
\caption{The structure and participating degrees of freedom of the steady state (s.s.) matrix in Eq.~\eqref{eq:parity} for different $\mathbb{Z}_2$ phases.} 
\label{tab:dof}
\end{table}

\textit{$\mathbb{Z}_2$-symmetric model.---}We  make this general analysis more concrete by focusing on a  minimal driven-dissipative example that exhibits both strong and weak versions of $\mathbb{Z}_2$ symmetry-breaking transitions in an infinite-size limit. Consider the rotating-frame Hamiltonian for a photonic cavity mode subject to a coherent two-photon drive:
\begin{equation} \label{eq:ham}
H = \omega a^\dagger a + \lambda \left( a^2 + (a^\dagger)^2 \right),
\end{equation}
where $\omega, \lambda \in \mathbb{R}$ \cite{cat-exp1, mirrahimi2014, ciuti-exact1, ciuti-exact2, roberts2020}.  The Hamiltonian possesses a $\mathbb{Z}_2$ symmetry with respect to Bose parity: $[H,P]=0$, where $P = \exp(i \pi a^\dagger a)$. Dissipation can be introduced in ways that respect strong or weak versions of the parity symmetry. We present our strong case along with the previously studied weak case \cite{ciuti2018}, further developing the latter.

In the strong case, we consider  two-photon loss $L_2=\sqrt{\kappa_2} a^2$ and dephasing $L_d= \sqrt{\kappa_d} a^\dagger a$. In the weak case, we add one-photon loss $L_1=\sqrt{\kappa_1}  a$ in addition to $L_2$ and $L_d$. Note: $[L_2, P]=[L_d, P]=0$ and $\{L_1, P\}=0$, which justifies our classification.  For both strong- and weak-symmetric dissipation, we expect a phase transition from an unbroken phase in the limit of small driving $ \lambda / \omega \ll 1$ to a broken phase in the limit of large driving $ \lambda / \omega \gg 1$, with a nonzero $\mathbb{Z}_2$-broken order parameter $\langle a \rangle$ in the steady state.

\begin{figure}
    \centering
    \includegraphics[scale=0.525]{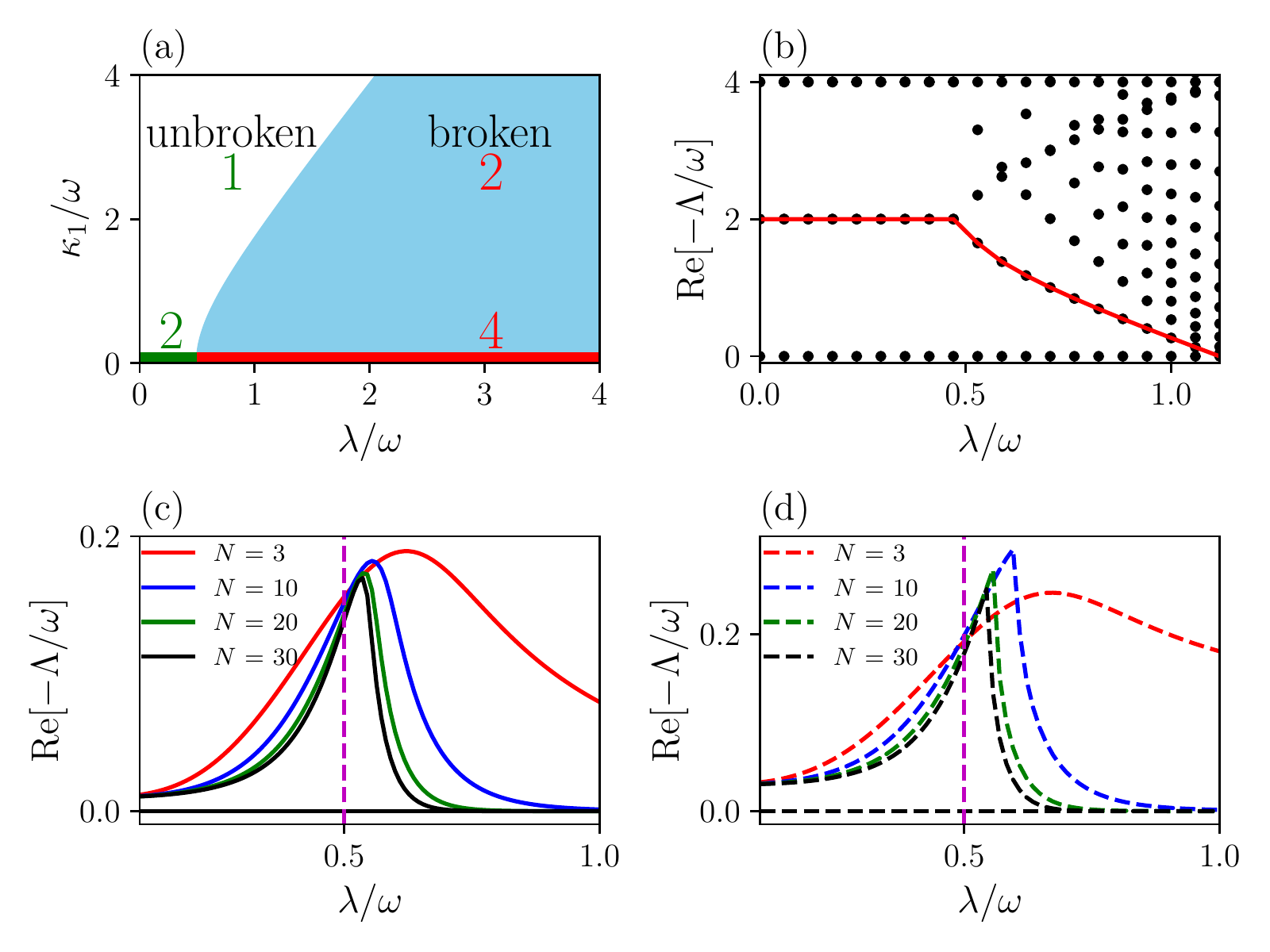}
    \caption{(a) Phase diagram for the model in Eq.~\eqref{eq:ham} with one-photon loss $\kappa_{1}$ in the limit $\kappa_2 = \kappa_d=0$. Integers indicate the dimension of the steady-state manifold. (b) Analytical expression for the dissipative gap (red line) and numerical spectrum (black dots) in the unbroken phase for $\kappa_2=\kappa_d=0, \kappa_1/\omega=2$. The dissipative gap closes as  the  phase boundary at $\lambda/\omega = \sqrt{5}/2 \approx 1.2$ is approached.    (c) Strong transition: Decay rate of the 4 modes with the longest lifetime; 2 modes are always pinned to zero and the other 2 are degenerate. A 2-to-4 transition occurs near $\lambda/\omega=0.5$ (purple dashed line) in the limit $N\rightarrow \infty$, in agreement with the phase diagram. $\kappa_1=0, \lambda/\kappa_2=N,  \kappa_d/\omega=0.01$. (d) Weak transition: Decay rate of the 2 modes with the longest lifetime. Dashed lines emphasize a lack of exact two-fold degeneracy present in (c). A 1-to-2 transition is observed. $\lambda/\kappa_2=N, \kappa_1/\omega=0.02, \kappa_d/\omega=0.01$.}
    \label{fig:mf-pd}
\end{figure}

We uncover the phase diagram  using two independent methods that agree: (1) a solution for the order parameter and (2) an  expression for the  dissipative gap. The expectation value of the order parameter $a$ satisfies
\begin{equation} \label{eq:mf}
\frac{d}{dt} \langle a \rangle =-2 i \lambda \langle a^\dagger \rangle - (i \omega + \kappa_1 + \kappa_d) \langle a \rangle -2 \kappa_2 \langle a^\dagger a^2 \rangle,
\end{equation}
where the right-hand-side follows from $\partial_t \langle a \rangle =  \text{Tr}[a \mathcal{L}(\rho)]$.  To determine the steady-state expectation value, we set $ \partial_t \langle a \rangle_{ss} =0$ and check which parameter regime produces non-trivial solutions for  $\langle a \rangle_{ss} \equiv \alpha$. In the mean-field approximation, $ \langle a^\dagger a^2 \rangle \approx |\alpha|^2 \alpha$, which is justified when $|\alpha|^2$ (the cavity photon population) is large. The critical boundary satisfies  $(\kappa_1 + \kappa_d) / \omega = \sqrt{4(\lambda / \omega)^2 -1}$, with a cavity photon population  $|\alpha|^2 =[\sqrt{4 \lambda^2 - \omega^2} - (\kappa_1 + \kappa_d)] /(2 \kappa_2)$  and $\text{arg}[\alpha] = \arccos[-\omega / (2 \lambda)]/2$ in the broken phase. The steady-state population of photons diverges as $\lambda/\kappa_2 \equiv N \rightarrow \infty$, which represents the thermodynamic limit for this model \cite{ciuti2018,carmichael2015, curtis2020, cirac2012}.   Fig.~\ref{fig:mf-pd}(a) presents the phase diagram for $\kappa_2=0$; the mean-field  equation is exact in this limit. Both weak ($\kappa_1 \neq 0$) and strong ($\kappa_1 = 0$) models indeed exhibit a transition characterized by  a $\mathbb{Z}_2$-broken order parameter $\langle a \rangle_{ss}$.

We  show that the dissipative gap closes at the critical boundary for $\kappa_2= \kappa_d=0$. In this (thermodynamic) limit,  $\mathcal{L}$ is quadratic in Bose operators, hence we can calculate the dissipative gap in the unbroken phase:  $\Delta_g = -\text{Re}[\kappa_1 +\sqrt{4 \lambda^2-\omega^2}]$ (see Supplemental Material (SM)  \cite{SM}). Setting $\Delta_g=0$ leads to a phase boundary which is identical to the mean-field analysis plotted in Fig.~\ref{fig:mf-pd}(a). Fig.~\ref{fig:mf-pd}(b) plots the expression for $\Delta_g$ along with a numerical calculation of the Lindblad spectrum $\{\Lambda\}$. We expect an extensive number of modes to touch zero at the critical point $\lambda \approx 1.1$, but our numerics are limited by a finite  Hilbert space. Similar results were recently reported in a related model \cite{baranger2020}.

Away from this exactly-solvable limit, i.e.~$\kappa_2 \neq 0$ and/or  $\kappa_d \neq 0$, we use numerical exact diagonalization to examine the steady-state dimension across the boundary.  Fig.~\ref{fig:mf-pd}(c) probes the strong transition 
by plotting the four spectral eigenvalues  with the smallest decay rate.  Indeed, two of these are always pinned to zero due to the strong symmetry, but two additional  zero eigenvalues appear in the broken phase. The transition occurs near values predicted by the phase diagram as the system approaches the thermodynamic limit $\lambda/\kappa_2 =  N \rightarrow \infty$. We repeat the analysis for the weak transition in Fig.~\ref{fig:mf-pd}(d) by plotting the two modes with the longest lifetimes and observe a 1-to-2 transition. This confirms our general analysis in Table \ref{tab:sym}.  The degeneracy at zero in the broken phase is split by an exponentially small term $\sim \exp(-N)$ (see SM \cite{SM}).

The rest of our analysis will focus on the strongly-symmetric model, setting $\kappa_1=0$. We inspect the nature of the steady states by writing down their exact expressions in extreme limits. First consider  the unbroken phase  $\omega \neq 0, \lambda =\kappa_d=0$. There are only two eigenoperators of $\mathcal{L}$ with zero eigenvalue  $| 0 \rangle \langle  0 |$ and $| 1 \rangle \langle  1 |$. The steady-state manifold reads  $\rho_{ss}(x) = x | 0 \rangle \langle  0 |  + (1-x)| 1 \rangle \langle  1 | $ for $x \in [0,1]$.  This represents a classical bit of information, since only relative magnitudes of an initial superposition are remembered, in agreement with Table \ref{tab:dof}.
 
Next consider the broken limit  $\omega = \kappa_d = 0, \lambda \neq 0$. Define the following coherent states  $
 \left| \pm \alpha \right\rangle = \sum_{n=0}^{\infty}   ( \pm \alpha )^n | n \rangle / \sqrt{n!}$ where $ \pm\alpha =  \pm e^{ i \pi/4}\sqrt{\lambda/\kappa_2}$. $\alpha$ matches the mean-field result, defined up to a minus sign degeneracy. Then  any pure state of the form $
\left| \psi \right\rangle = c_e\left| \alpha \right\rangle_e +c_o  \left| \alpha \right\rangle_o$ will be a steady state, where we  define normalized even and odd ``cat'' coherent states  $\left| \alpha \right\rangle_{e,o} \propto \left| \alpha \right\rangle  \pm \left| - \alpha \right\rangle  $ \cite{knight1994}.  An arbitrary superposition of these cat states is a steady state, an example of a decoherence-free subspace (DFS) \cite{lidar1998}.

\textit{Passive error correction for cat qubits.---}We now  show that a qubit encoded in the steady-state subspace of the strong-broken phase  benefits from passive error correction in the thermodyanamic limit $\lambda/\kappa_2 = N \rightarrow \infty$. We have just seen that  the limit  $\kappa_1 = \kappa_d= \omega=0$ hosts a DFS spanned by cat states. We define $\mathcal{L}_{0}$ to be the Lindbladian at this point. Previous studies have suggested that this coherent subspace could serve as a platform for universal quantum computation  that is intrinsically protected against dephasing errors \cite{mirrahimi2014}. Ref.~\cite{mirrahimi2014} found that, as $|\alpha|^2 \rightarrow \infty$,   an initially pure cat qubit, which encounters a dephasing  term in the Lindbladian for a short time (with respect to the inverse dissipative gap)  will return to its initial pure state after evolving the system with $\mathcal{L}_{0}$. In this context, our analysis allows us to: (1) extend the protection to errors that last an \textit{arbitrary} amount of time (cf. \cite{cohenthesis}), (2) understand the dynamics of the state throughout the error process, and  (3)  classify the types of errors that  self correct via the environment. This has direct experimental consequences for near-term quantum computing with photonic cat states \cite{cat-exp1, cat-exp2, cat-exp3, leghtas2020, touzard2018, grimm2019}. 

We consider the following protocol: Initialize the system in a pure state $\rho_i = |\psi \rangle \langle \psi |, |\psi \rangle = c_e  |\alpha\rangle_e +c_o  |\alpha\rangle_o $, which represents the qubit and satisfies $\mathcal{L}_{0}( \rho_i) =0$. Then quench the state with an ``error''  for an arbitrary time $\tau_q$ to obtain $\rho_m = \exp{[(\mathcal{L}_{0}+ \mathcal{L}' )\tau_q]} (\rho_i)$. Finally, turn off the error and evolve the system with $\mathcal{L}_0$ for a long time such that it reaches its steady state: $\rho_f = \lim_{t\rightarrow \infty} \exp{[\mathcal{L}_0 t]} (\rho_m) $. For what types of perturbations  $\mathcal{L}'$ will $\rho_f$ and $\rho_i$ be equal?

\begin{figure}[t]
    \centering
    \includegraphics[scale=0.525]{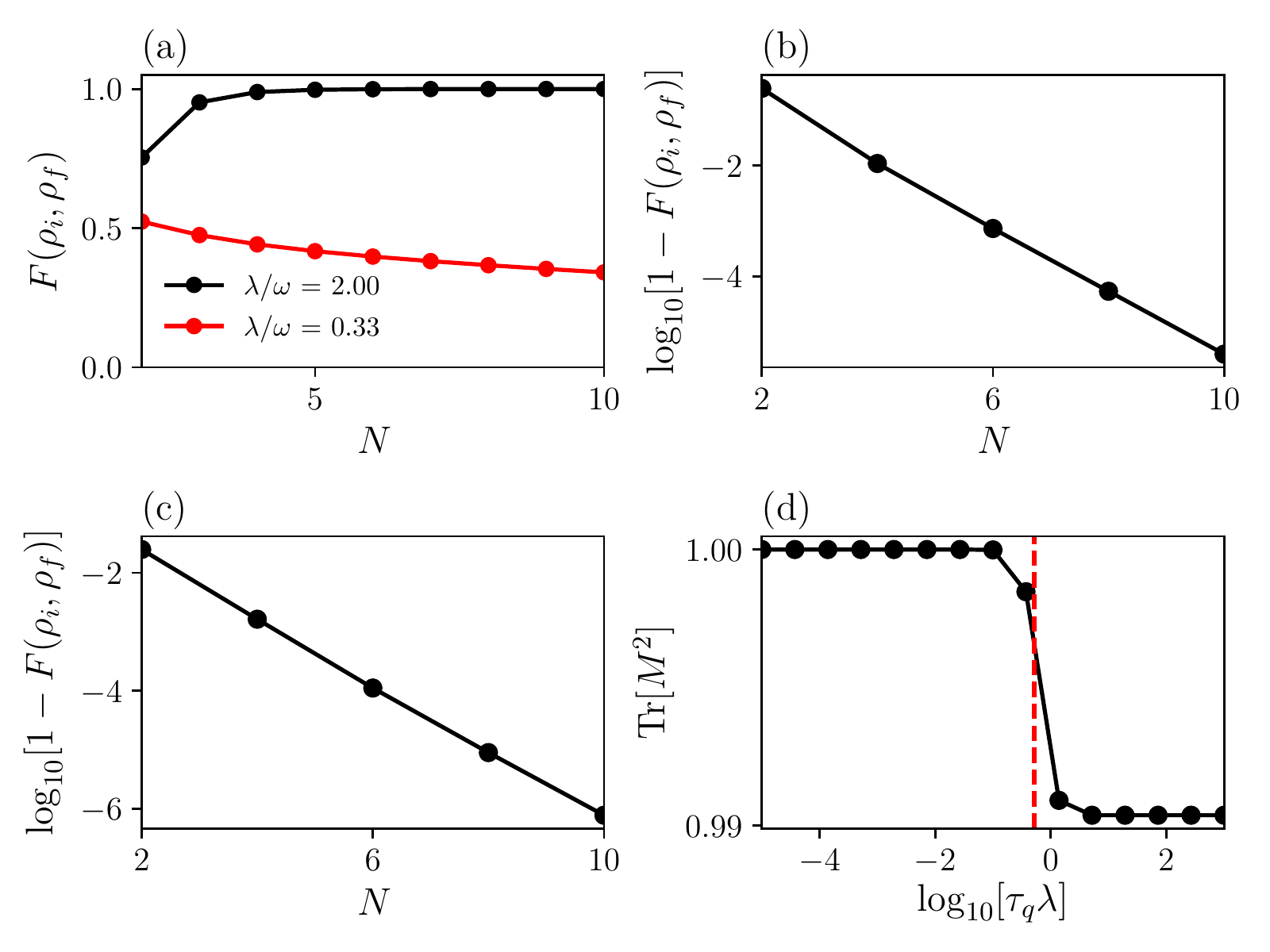}
    \caption{(a) Fidelity of the initial and final states for the quench protocol given in the main text with $\lambda/\kappa_2=N, \kappa_d=0, \tau_q \lambda=10$, $F(\rho_i, \rho_f) = \text{Tr}[\sqrt{\sqrt{\rho_i} \rho_f \sqrt{\rho_i}}]^2$.  Quenches to the strong-broken phase (black dots)  have a fidelity that tends to one in the thermodynamic limit, while quenches to the strong-unbroken phase (red dots) do not. (b) Same parameters as in (a) with $\lambda / \omega=2$; the fidelity tends to one exponentially fast in $N$. (c) A  dephasing error  $\kappa_d/ \lambda=0.03, \omega=0, \lambda/\kappa_2=N, \tau_q \lambda=10$; again the fidelity  is exponentially close to one. (d) Purity of $M$ [see Eq.~\eqref{eq:middle_struc_mt}] for different quench times with the same parameters as in (b) and $N=15$. Dashed line is the time scale set by the dissipative gap  $\tau_g= \Delta_g^{-1}$ of $\mathcal{L}_0+\mathcal{L}'$. Short quenches keep the system approximately pure, while long quenches evolve the system to a mixed NS steady state. Errors are correctable in both cases. For all figures,  $c_e=1/\sqrt{2}, c_o=i/\sqrt{2}$.  }
    \label{fig:f2}
\end{figure}

In Fig.~\ref{fig:f2}(a,b), we plot the fidelity $F$ between the initial state and the final state for the protocol described above with an error in the frequency, i.e.\ $H'=\omega a^\dagger a$, which either keeps the system  in the strong-broken phase (black dots) or moves it to the strong-unbroken phase (red dots).  The fidelity tends to one exponentially fast in cavity photon number for a long quench time $\tau_q$ \textit{only} if the perturbation kept the system in the broken phase. Fig.~\ref{fig:f2}(c) shows a similar behavior in the presence of a dephasing error: The qubit is able to perfectly correct itself as $N\rightarrow \infty$.

We can understand this striking behavior by recalling that the system is guaranteed to host a qubit steady state structure in the $N\rightarrow \infty$ limit  of the strong-broken phase. Away from the special point $\mathcal{L}_0$ but within the strong-broken phase, our numerics suggest that the steady-state structure is a noiseless subsystem (NS) \cite{knill2000}: a qubit in any  state tensored with a fixed mixed state. In other words, at any time after the introduction of the error, the state has the form
\begin{align} \label{eq:middle_struc_mt}
\rho_m(\tau_q)=  
\left(\begin{array}{cc}
|c_e|^2 & c_e  c_o ^*\\
c_e^* c_o  & |c_o |^2 \\
\end{array}\right) \otimes M(\tau_q)~,
\end{align}
where the qubit factor remains perfectly encoded in the even/odd parity basis, while the state $M(\tau_q)$ interpolates between the (pure) DFS steady state and the (mixed) NS steady state. The purity of $M(\tau_q)$ for different quench times is given in Fig.~\ref{fig:f2}(d), corroborating this interpretation: Short quenches  leave $M$ approximately pure, while long quenches allow it to equilibrate to a mixed steady state (cf.~\cite{passive7}). In both cases, the initial qubit state can be restored via evolution by $\mathcal{L}_0$. This decoupling of the qubit from auxiliary modes is reminiscent of the decoupling  used in quantum-information-preserving sympathetic cooling of trapped ions \cite{wang17c} and neutral atoms \cite{belyansky2019}, as well as in the nuclear-spin-preserving manipulation of electrons in alkaline-earth atoms \cite{reichenbach07,gorshkov2009}. The SM \cite{SM} provides numerical evidence for the structure in Eq.~\eqref{eq:middle_struc_mt}, including  the NS steady-state of $\mathcal{L}_0 + \mathcal{L'}$. The SM \cite{SM} also shows  perfect recovery of the fidelity for long quenches via  an independent method, i.e.~using asymptotic projections \cite{albert2016}.

\begin{table}
\begin{center}
\begin{tabular}{|c|c|c|c|}
\hline 
 error  & strong? & broken? & correcting? \tabularnewline
\Xhline{3\arrayrulewidth}
$L_1'=\sqrt{\kappa_1} a, \lambda/\kappa_1>0.5$ & no & yes & no \tabularnewline
\hline 
$H'=\omega a^\dagger a,  \lambda/\omega <0.5 $ & yes & no & no \tabularnewline\hline 
$H'=\omega a^\dagger a, \lambda/\omega >0.5$ & yes & yes & yes \tabularnewline
\hline 
$L_d'=\sqrt{\kappa_d} a^\dagger a, \lambda/\kappa_d>0.5$ & yes & yes & yes \tabularnewline
\hline 
\end{tabular}
\end{center}
\caption{ Examples of errors that can and cannot be passively corrected via evolution by $\mathcal{L}_{0}$ for the protocol given in the main text.  An error must preserve the strong symmetry \textit{and} keep the model in the broken phase in order for the final state to match the initial one.} 
\label{tab:errors}
\end{table}

The argument above relies on the presence of a qubit steady-state structure for $\mathcal{L}_{0}+ \mathcal{L}'$ in the large-$N$ limit. In its absence, the error will immediately cause the state to  lose information about the relative magnitude and/or phase of $c_e, c_o, $ which define the qubit. We conjecture that  any error  $\mathcal{L}'$ which keeps the model in the strong-broken phase can be passively corrected, which agrees with Fig.~\ref{fig:f2}(a). Table \ref{tab:errors} provides a list of potential errors. Our framework  allows us to classify  the terms that are expected to  self correct via the environment. Analytical proof of this conjecture requires an exact solution for the steady states (including mixed-parity sectors) in the entire strong-broken phase---an open direction for future work.

\textit{Summary and outlook.---} We uncover the distinction between strong and weak symmetry-breaking transitions in open systems and show that a qubit can be encoded into the steady state of the strong-broken phase. This qubit benefits from passive error correction: Any error induced on the qubit via a symmetric term that preserves the dissipative gap can be fixed by evolving with the environment in the thermodynamic limit.  

While we have studied a  $\mathbb{Z}_2$-symmetric system---the two-photon cat code---a  $\mathbb{Z}_n$-symmetric model should host a similarly protected qu$n$it in the strong-broken phase. Our symmetry-breaking analysis should also apply to related examples in Dicke-model physics \cite{kirton2019}, multi-mode systems \cite{paircat}, and  molecular platforms \cite{albert2019}. Finding protected qubits and strong symmetry-breaking transitions in models with a local finite-dimensional Hilbert space (e.g.~a driven-dissipative Ising model \cite{keeling2013, rossini2018}) remains an interesting question for future work.

Our predictions regarding qubit stability in the strong-broken phase should be observable using available experimental setups. Cat qubits of light encoded in superconducting resonators with dominant two-photon loss channels have enjoyed recent success \cite{cat-exp1, cat-exp2, cat-exp3, leghtas2020, touzard2018, grimm2019}. It would be interesting to perform the ($\omega$) quench protocol outlined in this paper (e.g.~by quenching the pump-cavity detuning). The driven-dissipative transition can then be determined by probing qubit fidelity. Our predictions can also be tested by engineering two-phonon loss \cite{zoller1996} and two-phonon drive \cite{burd19} for a motional mode of a trapped ion. 

Strong dissipative transitions may also represent a fundamentally new class of non-equilibrium criticality. To our knowledge, all previous studies of non-equilibrium transitions fall into the category of weak symmetry breaking. An important  question remains to examine the critical exponents of strong transitions \footnote{R.~Belyansky \textit{et al}, in preparation.}.

In  closed quantum systems, symmetry-breaking transitions can be dual to topological transitions. For example, the  transverse-field Ising model undergoes a symmetry-breaking transition that maps  to a topological transition in a fermionic (Kitaev) chain \cite{kitaev2001}.  Recent efforts have generalized different aspects of topological matter to open systems \cite{bergholtz2019,  altland2020, wang2019, chen2020,yoshida2020, nori2020}; in particular, zero-frequency edge modes with a finite lifetime can be protected via a \textit{frequency} gap \cite{lieu2020}. An open question remains whether a dissipative topological  phase can be characterized by edge modes with zero decay rate. The resulting qubit steady state structure would be  immune to all local error channels that preserve the \textit{dissipative} gap. Such a model remains elusive,  representing an exciting avenue for future research.

\begin{acknowledgments}
S.L.\ was supported by the NIST NRC Research Postdoctoral Associateship Award. R.B., J.T.Y., R.L., and A.V.G.\ acknowledge funding by the DoE ASCR Accelerated Research in Quantum Computing program (award No.~DE-SC0020312), NSF PFCQC program, DoE BES Materials and Chemical Sciences Research for Quantum Information Science program (award No.~DE-SC0019449), DoE ASCR Quantum Testbed Pathfinder program (award No.~DE-SC0019040), AFOSR, AFOSR MURI, ARO MURI, ARL CDQI,  and NSF PFC at JQI.
R.B. acknowledges support of NSERC and FRQNT of Canada. 
\end{acknowledgments}

\bibliography{sw,vva}
\bibliographystyle{apsrev4-1}

\newpage
\afterpage{\blankpage}

\newpage
\widetext

\setcounter{equation}{0}
\setcounter{figure}{0}
\setcounter{table}{0}
\setcounter{page}{1}

\renewcommand{\theequation}{S\arabic{equation}}
\renewcommand{\thefigure}{S\arabic{figure}}

\begin{center}
		{\fontsize{12}{12}\selectfont
			\textbf{Supplemental Material for  ``\papertitle''\\[5mm]}}
		{\normalsize \authornames\\[1mm]}
		{\fontsize{9}{9}\selectfont  
			$^1$\textit{Joint Quantum Institute, NIST/University of Maryland, College Park, MD 20742, USA} \\			
			$^2$\textit{Joint Center for Quantum Information and Computer Science, \\ NIST/University of Maryland, College Park, Maryland 20742 USA}\\			
			$^3$\textit{Institute for Quantum Information and Matter and Walter Burke Institute for \\ Theoretical Physics,
California Institute of Technology, Pasadena, CA 91125, USA}}
\end{center}
\normalsize\

In Sec.~1, we analytically show that the dissipative gap closes at the critical point by utilizing an exact solution for the Lindblad spectrum [Fig.~\ref{fig:mf-pd}(b) in the main text].  Sec.~2 exhibits  numerical evidence for a noiseless subsystem steady state  in the strong-broken phase (away from $\mathcal{L}_0$). Sec.~3 tracks the evolution of the state throughout the error protocol in the main text. We show numerical evidence for the state structure defined in Eq.~\eqref{eq:middle_struc_mt} of the main text for errors which keep the model in the strong-broken phase. Sec.~4 uses the asymptotic projection method to confirm perfect fidelity recovery in the thermodynamic limit, in agreement with the direct numerical evolution discussed in the main text.

\section{ 1.~Closing of the dissipative gap at the critical point} \label{sec:prosen}

\begin{figure}[b]
    \centering
    \includegraphics[scale=0.35]{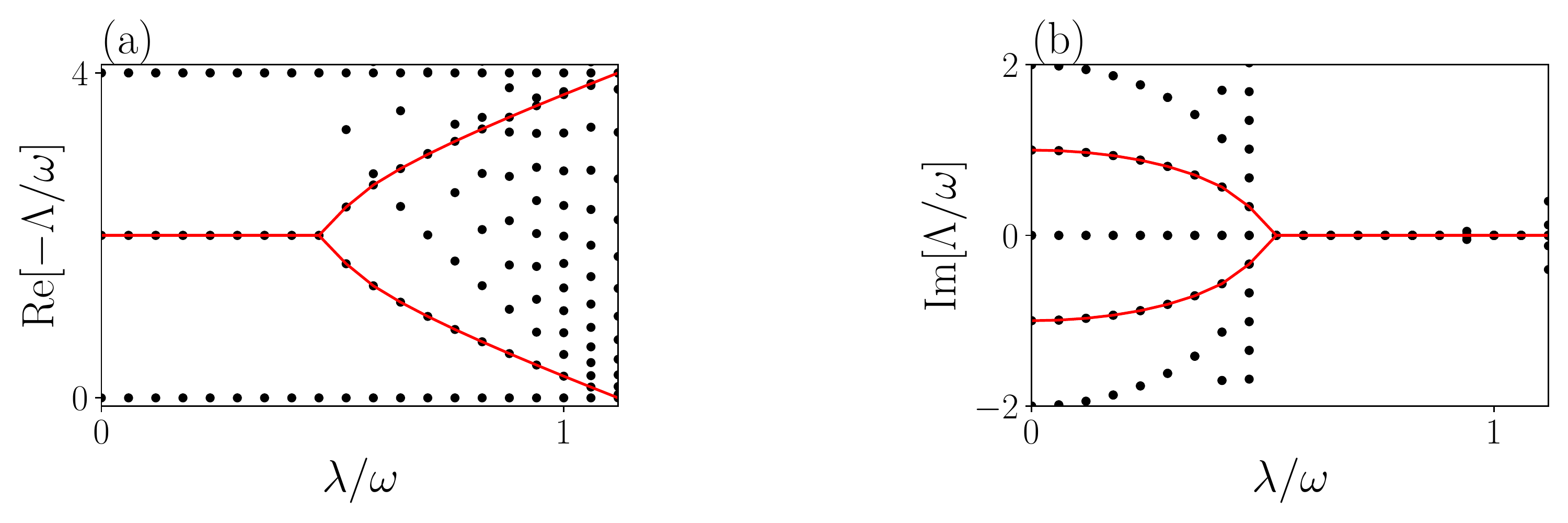} 
    \caption{ Analytical single-particle spectrum (red lines) and numerical many-body spectrum (black dots) with $\kappa_1/\omega = 2, \kappa_2=\kappa_d=0$. The many-body spectrum comes in integer multiples of the single-particle excitations. As the system approaches the  critical point from the unbroken phase, the single-particle spectrum touches zero at the phase boundary  $\lambda /\omega  \approx 1.1$.  The numerical spectrum starts to deviate from the analytical predictions near the transition  due to truncation of the Hilbert space dimension $d_\text{Hilbert}=70$. We plot up to  25 eigenvalues closest to zero for clarity.   }
    \label{fig:sp-spec}
\end{figure}

We show that an extensive number of spectral eigenvalues touch zero at the critical boundary  [Fig.~\ref{fig:mf-pd}(a) in the main text] when approaching from the unbroken phase in the thermodynamic limit. We utilize Prosen's ``third quantization'' technique which allows us to fully diagonalize a quadratic Lindbladian \cite{prosen2008, prosen2010}. For the Hamiltonian \eqref{eq:ham} in the presence of  one-photon loss only (i.e.~the weak transition), the Lindbladian can be expressed as $
\mathcal{L} = \epsilon_+ \beta^\dagger_+ \beta_+' + \epsilon_- \beta^\dagger_- \beta_-',$ where $\beta$ are bosonic superoperators satisfying generalized commutation relations  $[\beta_{i}', \beta_{j}^\dagger] = \delta_{ij}$. These excite a quantum of ``complex energy''  $\epsilon_\pm = -\kappa_1 \pm \sqrt{4 \lambda^2 - \omega^2}$, where the (unique) steady state is annihilated by all quasiparticles  $\beta_{\pm}' \rho_{ss}=0$, and the  many-body spectrum is built from these single-particle excitations  $\mathcal{L}  [(\beta_{+}^\dagger)^n (\beta_{-}^\dagger)^m \rho_{ss}] = (n \epsilon_+ + m \epsilon_-) [(\beta_{+}^\dagger)^n (\beta_{-}^\dagger)^m \rho_{ss}]$. The single-particle spectrum touches zero  at $\kappa_1 / \omega = \sqrt{4(\lambda / \omega)^2-1}$, which coincides with the emergence of a non-zero order parameter (see main text).  This implies that an infinite number of eigenvalues of $\mathcal{L}$ are zero at the critical point of the weak transition from 1 steady state to 2 steady states. We plot both the single-particle spectrum  and match it  with many-body numerics  in Fig.~\ref{fig:sp-spec}. [Fig.~\ref{fig:sp-spec}(a) and Fig.~\ref{fig:mf-pd}(b) are equivalent; here we plot the real and imaginary parts side by side.] The numerical spectrum deviates from analytical predictions only near the critical boundary due to  truncation of the Hilbert space dimension. Note that the analytical and numerical plots are only valid in the unbroken phase. The steady state has an infinite number of photons  in the broken phase, hence any finite-size Hilbert space will not produce a converged spectrum. Finite-size scaling [Fig.~\ref{fig:mf-pd}(d)] suggests that two eigenvalues are exponentially close to zero in the weak-broken phase with a dissipative gap to the rest of the modes. 

\section{ 2.~Noiseless subsystem in the strong-broken phase} \label{sec:ns}

We demonstrate that the model described in the main text possesses a qubit steady-state structure in the thermodynamic limit of the strong-broken phase. In particular, we will show that  the four  right eigenoperators with zero eigenvalue  can be written in the form  $r_{\mu \nu} = \left| \mu \right \rangle \left\langle \nu \right| \otimes z$ with $(\mu, \nu) \in (+,-)$. This is called a noiseless subsystem (NS) if $z$ is mixed, and a decoherence-free subspace (DFS) if $z$ is pure \cite{albert2014, knill2000, lidar1998}.

The four steady-state right eigenoperators  belonging to the different parity sectors are
\begin{equation} \label{eq:rhos}
r_{++}^F = \left(\begin{array}{cc}
s_{++} & 0\\
0 & 0
\end{array}\right),\qquad
r_{--}^F = \left(\begin{array}{cc}
0 & 0\\
0 & s_{--}
\end{array}\right),\qquad
r_{+-}^F = \left(\begin{array}{cc}
0 & s_{+-}\\
0 & 0
\end{array}\right),\qquad
r_{-+}^F = \left(\begin{array}{cc}
0& 0\\
s_{-+} & 0
\end{array}\right)
\end{equation}
in the Fock basis  $[ \ket{0},\ket{2}, \ket{4},\ldots,  \ket{1},\ket{3}, \ket{5},\ldots ]^T$. They each satisfy $\mathcal{L} (r)=0$ (in the thermodynamic limit). Since $s_{++}, s_{--}$ are guaranteed to be  Hermitian matrices, we can diagonalize them via a unitary transformation  $U = \text{Diag}[U_+, U_-]$ which relates the Fock basis to the diagonal basis $r_i^d = U^\dagger r_i^F U$. In this new basis, the eigenoperators are
\begin{equation}
r_{++}^d = \left(\begin{array}{cc}
z_{++} & 0\\
0 & 0
\end{array}\right),\qquad
r_{--}^d = \left(\begin{array}{cc}
0 & 0\\
0 & z_{--}
\end{array}\right),\qquad
r_{+-}^d = \left(\begin{array}{cc}
0 & z_{+-}\\
0 & 0
\end{array}\right),\qquad
r_{-+}^d = \left(\begin{array}{cc}
0& 0\\
z_{-+} & 0
\end{array}\right),
\end{equation}
where $z_{++},z_{--}$ are diagonal by construction, and $z_{+-},z_{-+}$ are diagonal in the thermodynamic limit. We will show that $z_{++}=z_{--}=z_{+-}=z_{-+}$ in this limit, which  implies that the system hosts  a NS or a DFS.

In the special limit  $\omega=\kappa_d=\kappa_1=0, \lambda\neq0, \kappa_2\neq0$, any pure superposition of even and odd cat states remains steady, as discussed in the main text. Thus  $z_{++}=z_{--}=z_{+-}=z_{-+}=\text{Diag}[1,0,0,0,\ldots]$, which implies a DFS.

\begin{figure}[b]
    \centering
    \includegraphics[scale=0.4]{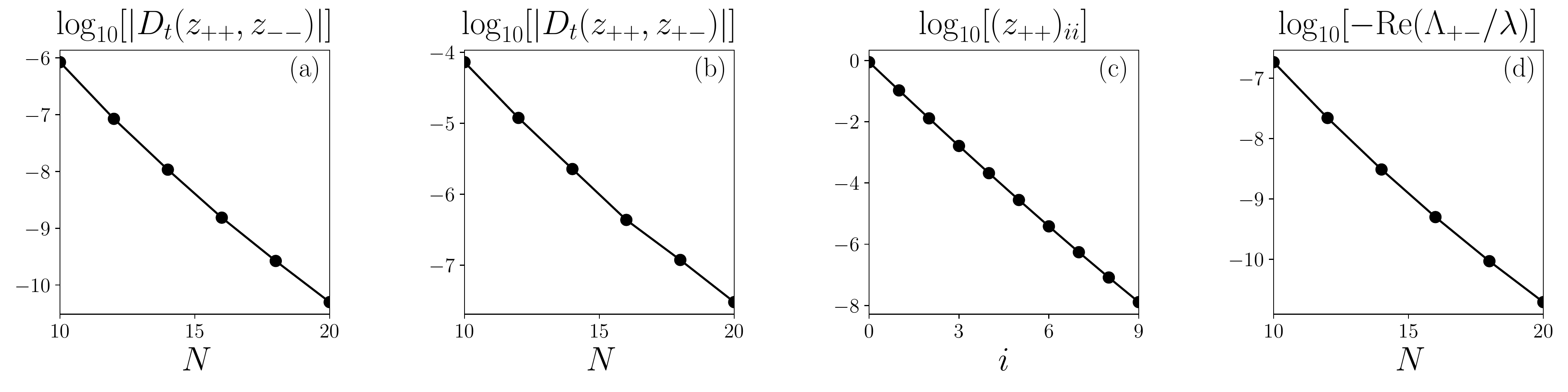} 
    \caption{ Parameters: $\lambda/\kappa_2=N, \kappa_d/\lambda=0.03, \omega=\kappa_1=0$, i.e.~non-zero dephasing. (a,b) The trace norm $D_t(A,B)=\text{Tr}[\sqrt{(A-B)^2}]/2$ between the different right eigenoperators with zero eigenvalue goes to zero in the thermodynamic limit $N\rightarrow \infty$. (c) Diagonal matrix elements of $z_{++}$ for $N=20$. The matrix is not pure, with elements scaling as $(z_{++})_{ii} \sim \exp{[-c i]}$ for some $c>0$. (d) The off-diagonal symmetry sector of the Lindbladian acquires an eigenvalue of zero as $N \rightarrow \infty$. Here $\Lambda_{+-}$ is the smallest eigenvalue in the off-diagonal sector.   }
    \label{fig:diff}
\end{figure}

We now consider a parameter regime away from this limit but within the strong-broken phase. We start by adding dephasing:   $\omega=\kappa_1=0,\kappa_d\neq0, \lambda\neq0, \kappa_2\neq0$. We will  numerically show that the $z$ matrices are equal and not pure. For the matrix distance, we choose the trace distance  $D_t(A,B)=\text{Tr}[\sqrt{(A-B)^2}]/2$. In Fig.~\ref{fig:diff}(a,b), we plot $D_t(z_{++}, z_{--})$ and $D_t(z_{++}, z_{+-})$ as the system approaches the thermodynamic limit $\lambda/\kappa_2=N\rightarrow \infty$. Indeed, we find that the matrices $z_{++}, z_{--}, z_{+-}$ all converge to a single matrix as $N$ is increased. ($z_{+-}$ and $z_{-+}$ are related by Hermiticity.) Additionally, in Fig.~\ref{fig:diff}(c), we show that $z_{++}$ is a non-pure matrix with elements that fall off as $(z_{++})_{ii} \sim \exp{[-i]}$. The purity of $z_{++}$ \textit{degrades} with $N$ (not shown). We conclude that  the system tends to a noiseless subsystem in the thermodynamic limit, since the $z_{\pm \pm}$ all converge to a single non-pure matrix. [For completeness, in Fig.~\ref{fig:diff}(d), we show that the smallest eigenvalue in the off-diagonal sector indeed tends to zero exponentially quickly with $N$. The steady-state degeneracy is split by an exponentially small factor, characteristic of symmetry-breaking transitions.]

\begin{figure}
    \centering
    \includegraphics[scale=0.4]{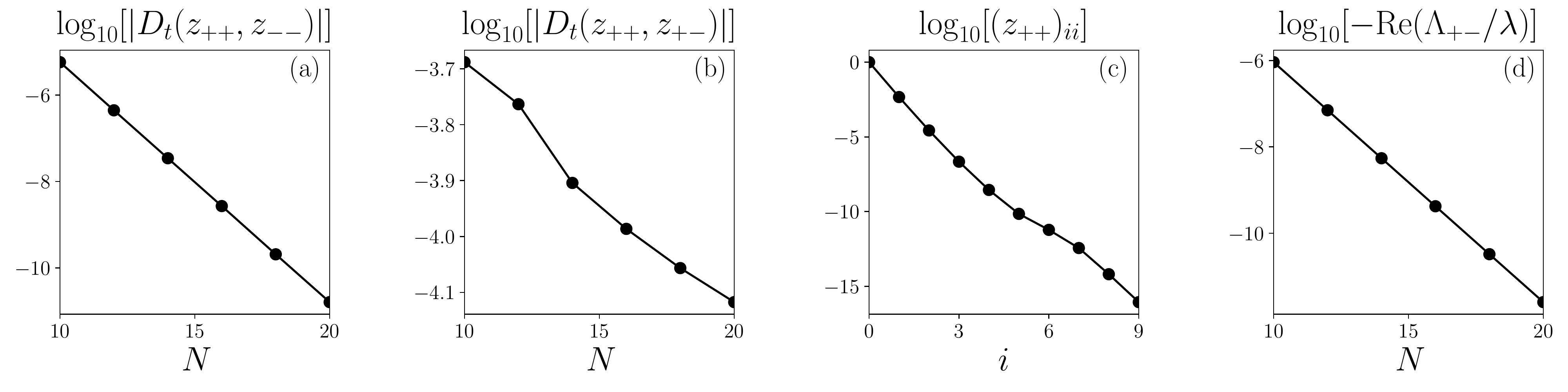} 
    \caption{ Parameters: $\lambda/\kappa_2=N, \omega/\lambda=0.5, \kappa_d=\kappa_1=0$. (a,b) The trace norm between the different right eigenoperators with zero eigenvalue goes to zero in the thermodynamic limit $N\rightarrow \infty$.  (c) Diagonal matrix elements of $z_{++}$ for $N=20$. The matrix is not pure, with elements scaling as $(z_{++})_{ii} \sim \exp{[-c i]}$ for some $c>0$. (d) The off-diagonal symmetry sector of the Lindbladian acquires an eigenvalue of zero as $N \rightarrow \infty$. Here $\Lambda_{+-}$ is the smallest eigenvalue in the off-diagonal sector.}
    \label{fig:diffom}
\end{figure}

We   repeat this analysis in the limit of no dephasing but non-zero $\omega$:   $\kappa_d=\kappa_1=0,\omega\neq0, \lambda\neq0, \kappa_2\neq0$. Fig.~\ref{fig:diffom} shows that the $z_{\pm \pm}$ converge to a single non-pure matrix  in the thermodynamic limit, similar to the case of dephasing. We therefore conclude that a generic model in the strong-broken phase possesses a noiseless subsystem, whilst a decoherence-free subspace exists at a special point $\mathcal{L}_0$ in the phase diagram.

\section{ 3.~Evolution from decoherence-free subspace to noiseless subsystem} \label{sec:dfsns}

We  now track the state throughout the error protocol described in the main text for both dephasing errors and Hamiltonian-frequency errors. Our analysis will confirm that the state can be written as a qubit tensored with a mixed state thoughout the entire quench protocol, i.e.~the structure described in Eq.~\eqref{eq:middle_struc_mt} in the main text.
 
We prepare the system in a pure steady state of $\mathcal{L}_{0}$:
\begin{equation}
\rho_i= 
\left(\begin{array}{cc}
|c_e|^2 & c_e c_o^*\\
c_e^* c_o & |c_o|^2 \\
\end{array}\right) 
\end{equation}
in the basis of even and odd cat states $\left| \alpha \right\rangle_e , \left| \alpha \right\rangle_o, $ where $|c_e|^2+|c_o|^2=1$ and  $\mathcal{L}_{0} (\rho_i) = 0$.  We evolve this initial state with an error  to a ``middle'' state
\begin{equation} \label{eq:rhom1}
\rho_m(\tau_q)= e^{(\mathcal{L}_{0}+\mathcal{L}') \tau_q} \rho_i.
\end{equation}
We wish to show that this middle state can be written in the form
\begin{equation} \label{eq:middle_struc}
\rho_m(\tau_q)= 
\left(\begin{array}{cc}
|c_e|^2 & c_e c_o^*\\
c_e^* c_o & |c_o|^2 \\
\end{array}\right)  \otimes M
\end{equation}
for some $M$ which is not necessarily pure.

We numerically solve for $\rho_m(\tau_q)$ via Eq.~\eqref{eq:rhom1} for arbitrary quench times  and $\mathcal{L}_{0}+\mathcal{L}'$ in the  strong-broken phase. We then split the matrix up into symmetry sectors in the Fock basis  $\rho_m = |c_e|^2 \rho_{++}^F + |c_o|^2 \rho_{--}^F + (c_e c_o^* \rho_{+-}^F + h.c.)$. The four operators  belonging to the different parity sectors are
\begin{equation} 
\rho_{++}^F = \left(\begin{array}{cc}
x_{++} & 0\\
0 & 0
\end{array}\right),\qquad
\rho_{--}^F = \left(\begin{array}{cc}
0 & 0\\
0 & x_{--}
\end{array}\right),\qquad
\rho_{+-}^F = \left(\begin{array}{cc}
0 & x_{+-}\\
0 & 0
\end{array}\right),\qquad
\rho_{-+}^F = \left(\begin{array}{cc}
0& 0\\
x_{-+} & 0
\end{array}\right)
\end{equation}
in the Fock basis  $[ \ket{0},\ket{2}, \ket{4},\ldots,  \ket{1},\ket{3}, \ket{5},\ldots ]^T$. Since $x_{++}, x_{--}$ are guaranteed to be  Hermitian matrices, we can diagonalize them via a unitary transformation  $V = \text{Diag}[V_+, V_-]$  which relates the Fock basis to the diagonal basis $\rho_i^d = V^\dagger \rho_i^F V$. In this new basis, the eigenoperators are 
\begin{equation}
\rho_{++}^d = \left(\begin{array}{cc}
M_{++} & 0\\
0 & 0
\end{array}\right),\qquad
\rho_{--}^d = \left(\begin{array}{cc}
0 & 0\\
0 & M_{--}
\end{array}\right),\qquad
\rho_{+-}^d = \left(\begin{array}{cc}
0 & M_{+-}\\
0 & 0
\end{array}\right),\qquad
\rho_{-+}^d = \left(\begin{array}{cc}
0& 0\\
M_{-+} & 0
\end{array}\right),
\end{equation}
where  all the $M$s are diagonal by construction. We now show that all $M$s converge to a single matrix in the thermodynamic limit, confirming the  form of Eq.~\eqref{eq:middle_struc}.

\begin{figure}
    \centering
    \includegraphics[scale=0.4]{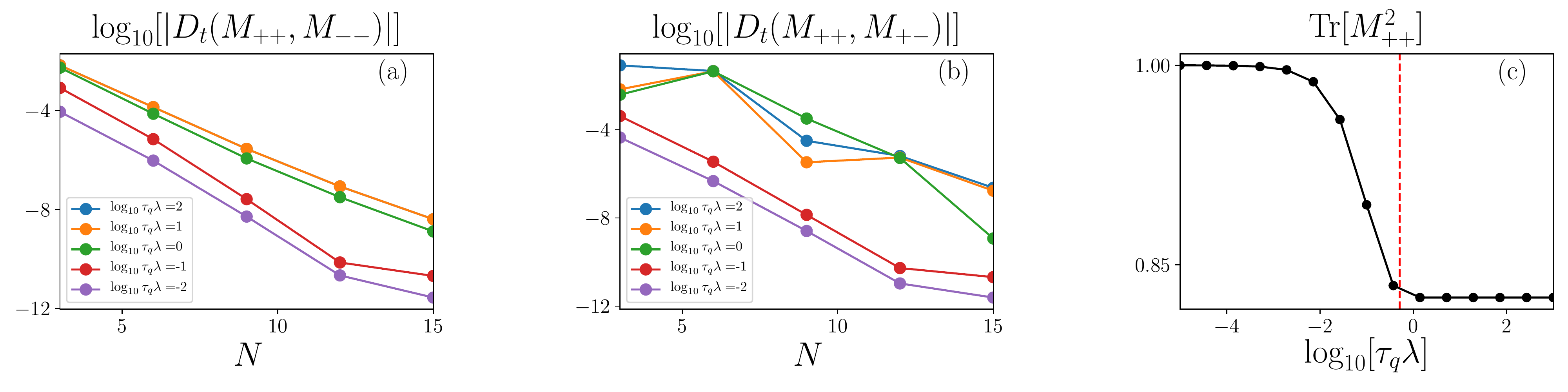} 

    \caption{ Parameters: $\lambda / \kappa_2 = N, \kappa_d / \lambda=0.03, \omega=\kappa_1=0, c_e=1/\sqrt{2}, c_o = i / \sqrt{2}$.  (a) The trace distance between $M_{++}$ and $M_{--}$ goes to zero exponentially fast in $N$. (b) Analogous behavior is observed for $M_{++}$ and $M_{+-}$.   (c) $N=15,$ the red line is the time scale set by the inverse dissipative gap $\tau_g = \Delta_g^{-1}$ of $\mathcal{L}_0 + \mathcal{L}'$. The state is approximately pure for short quenches compared to this time scale, while it settles to its (mixed) steady-state value for quenches longer than this timescale. }
    \label{fig:depq}
\end{figure}

\begin{figure}
    \centering
    \includegraphics[scale=0.4]{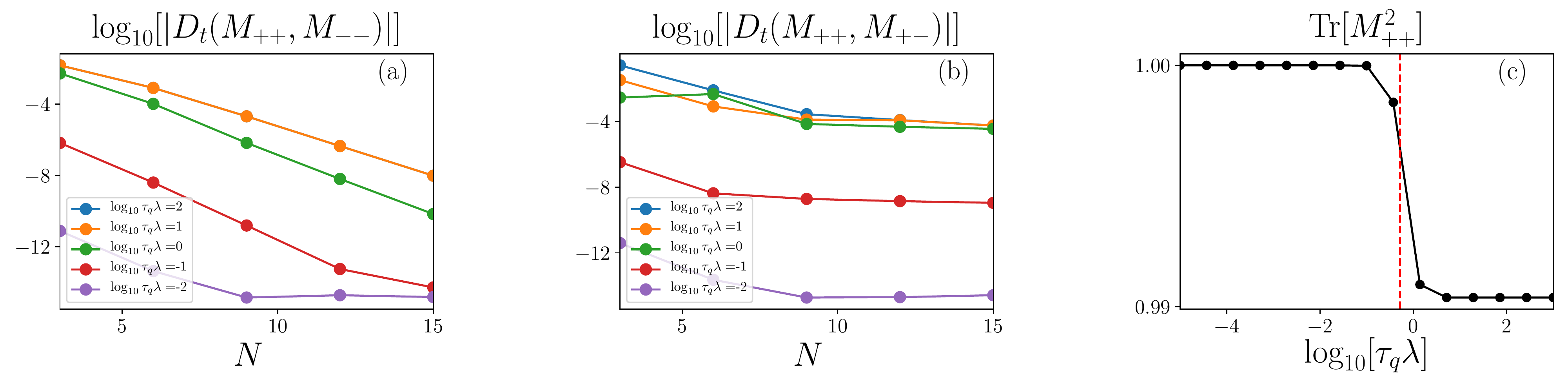}
    \caption{ Parameters: $\lambda / \kappa_2 = N, \omega/ \lambda=0.5, \kappa_d=\kappa_1=0, c_e=1/\sqrt{2}, c_o = i / \sqrt{2}$. (a) The trace distance between $M_{++}$ and $M_{--}$ goes to zero exponentially fast in $N$. (b) Analogous behavior is observed for $M_{++}$ and $M_{+-}$. (c) $N=15,$ the red line is the time scale set by the inverse dissipative gap $\tau_g = \Delta_g^{-1}$ of $\mathcal{L}_0 + \mathcal{L}'$. The state is approximately pure for short quenches compared to this time scale, while it settles to its (mixed) steady-state value for quenches longer than this timescale.  }
    \label{fig:omq}
\end{figure}

We plot the trace distance between the different $M$s for both short and long quench times $\tau_q \lambda \in [10^{-2} ,10^2]$.  In Fig.~\ref{fig:depq}, we consider a quench in the dephasing strength. Indeed, the trace distance between the different $M$s goes to zero exponentially fast as a function of $N$,  which suggests that the ansatz in Eq.~\eqref{eq:middle_struc} is correct in the limit $N \rightarrow \infty$.  We also track the purity of this matrix: At quench times that are short  compared to the timescale set by the dissipative gap (red line), the middle state remains approximately pure, whilst longer quenches imply that the system settles into its new steady state, which is  mixed (see previous section).  Analogous behavior is observed for a quench in frequency (Fig.~\ref{fig:omq}).

\section{ 4.~Asymptotic projection} \label{sec:ap}

We verify the perfect recovery of the fidelity observed in Fig.~\ref{fig:f2} of the main text via the asymptotic projection method \cite{albert2016}. Fig.~\ref{fig:f2} shows that qubit cat states will self correct via the environment if $\mathcal{L}_{0} + \mathcal{L}'$ remains in the strong symmetry-broken phase. This behavior can be understood via  perturbation theory for short quenches (compared to the time scale set by the dissipative gap) \cite{mirrahimi2014}. Here, we consider long quench times where the system evolves into the steady state of $\mathcal{L}_{0} + \mathcal{L}'$. Remarkably, such a drastic error can still be passively corrected via the environment $\mathcal{L}_{0}$. We provide simple expressions relating the initial, intermediate,  and final states by projecting onto the corresponding steady state manifolds.

Defining our initial state as $\rho_i$, we  evolve it with an error $(\mathcal{L}_{0}+\mathcal{L}')$ to a ``middle'' state  $\rho_m(\tau_q)= e^{(\mathcal{L}_{0}+\mathcal{L}') \tau_q} \rho_i$. We then evolve the state with $\mathcal{L}_{0}$ for an infinite time to reach the final state  $\rho_f(\tau_q) = \lim_{ \tau \rightarrow \infty} e^{\mathcal{L}_{0} \tau} \rho_m(\tau_q).$ We will discuss how $\rho_{i,m,f}$ relate to one another in this protocol when $\tau_q$ is much longer than the inverse dissipative gap of $\mathcal{L}_0 + \mathcal{L}'$. 

We first prepare the system in a pure steady state of $\mathcal{L}_{0}$,
\begin{equation}
\rho_i = |a|^2  r_{++}^0 + |b|^2  r_{--}^0 + a^* b r_{+-}^0 + a b^* r_{-+}^0,
\end{equation}
where $r_{++}^0 = \left| \alpha \right\rangle_e \left\langle \alpha \right|_e$, $r_{--}^0 = \left| \alpha \right\rangle_o \left\langle \alpha \right|_o$, $r_{+-}^0 = \left| \alpha \right\rangle_e \left\langle \alpha \right|_o$, $r_{-+}^0 = \left| \alpha \right\rangle_o \left\langle \alpha \right|_e$; $ \left| \alpha \right\rangle_{e/o} $ is the even/odd cat state, and  $\mathcal{L}_{0} (r^0_{\pm \pm}) = 0$. To find $\rho_m$, it is useful to define the right and left eigenoperators of the error:
\begin{equation}
(\mathcal{L}_{0} +\mathcal{L}')  (\tilde{r}_j) = \tilde{\Lambda}_j  (\tilde{r}_j), \qquad (\mathcal{L}_{0}^\dagger +\mathcal{L}'^\dagger)  (\tilde{l}_j) = \tilde{\Lambda}_j^* (\tilde{l}_j),
\end{equation}
where the spectrum $\{ \tilde{\Lambda} \}$ and eigenoperators determine the dynamics under $\mathcal{L}_{0} +\mathcal{L}'$. Assuming that the error keeps the system in the strong-broken phase, we know that two eigenvalues will be exactly zero  $\tilde{\Lambda}_{++}^0 = \tilde{\Lambda}_{--}^0 =0$ and two eigenvalues will be exponentially close to zero  $\tilde{\Lambda}_{+-}^0 = (\tilde{\Lambda}_{-+}^0)^* \sim e^{-N}$. We label the eigenvalue of the first ``excited'' state (above these four) as $\tilde{\Lambda}_g$, which sets the dissipative gap  in the thermodynamic limit. The exact expression for $\rho_m(\tau_q)$ reads 
\begin{equation} \label{eq:rhomexact}
\rho_m(\tau_q)= \sum_j   \exp[\tilde{\Lambda}_j \tau_q]  \text{Tr}[\tilde{l}_j^\dagger \rho_i ] \tilde{r}_j,
\end{equation}
where we have used the orthogonality relation  $\text{Tr}[\tilde{l}_j^\dagger \tilde{r}_k] = \delta_{jk}$. $-\text{Re}[\tilde{\Lambda}_j^{-1}]$ sets the lifetime of each eigenoperator. Consider  a quench time that obeys  $-\text{Re}[\tilde{\Lambda}_g^{-1}] \ll \tau_q \ll -\text{Re}[(\tilde{\Lambda}_{+-}^0)^{-1}] \sim e^N$. This quench is long enough for the system to relax into the new steady state but not so long that coherences are lost.  In this regime, $\rho_m$ will tend to the following matrix $t_m$ 
\begin{equation} \label{eq:rhom}
\lim_{N \rightarrow \infty}  \rho_m(\tau_q) = t_m, \qquad t_m= |a|^2  \tilde{r}_{++}^0 + |b|^2  \tilde{r}_{--}^0 + \left[a^* b \gamma_m  \tilde{r}_{+-}^0 + h.c. \right], \qquad \gamma_m =  \text{Tr}  \left[ (\tilde{l}_{+-}^{0})^\dagger r_{+-}^0  \right].
\end{equation}
If $\tau_q$ is longer than $-\text{Re}[\tilde{\Lambda}_g^{-1}]$,  then all excitations will vanish and we will be left with the projection onto the steady-state manifold of the error. We have confirmed this numerically  by doing the full time evolution $\rho_m = \exp{[(\mathcal{L}+\mathcal{L}') \tau_q]}\rho_i$ and comparing the resulting matrix with $t_m$. Indeed, the trace distance $D_t(\rho_m,t_m)=\text{Tr}[\sqrt{(\rho_m-t_m)^2}]/2$) goes to zero exponentially quickly in $N$.  We have thus found a simple expression for $\rho_m(\tau_q) $ for this range of $\tau_q$.

\begin{figure}[!]
    \centering
    \includegraphics[scale=0.3]{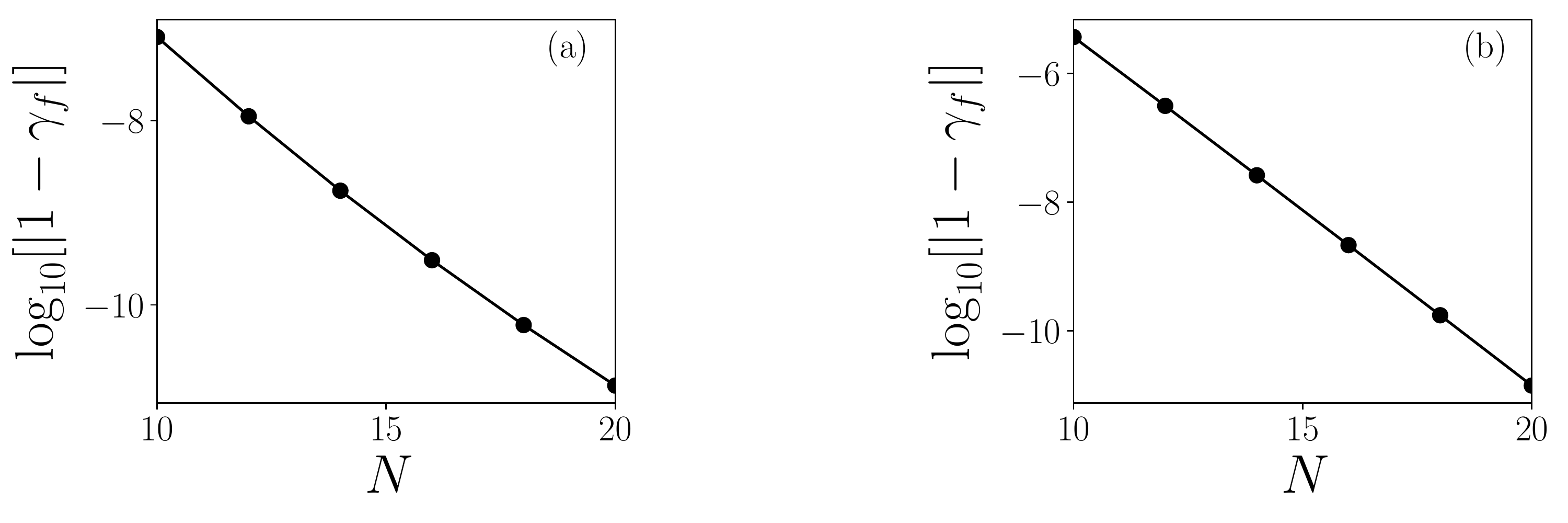}
    \caption{ Scaling of $|1-\gamma_f|$ as a function of $N$ for (a) a dephasing error  $ \lambda / \kappa_2 = N, \kappa_d/ \lambda=0.03$ and (b) a frequency error  $\lambda / \kappa_2 = N, \omega/ \lambda=0.03$. $\gamma_f$ approaches one exponentially fast in $N$ for both cases. }
    \label{fig:gamma-abs}
\end{figure}

Having understood the structure of this intermediate state, $\rho_m \approx t_m$, we now project this state back onto the steady-state manifold of $\mathcal{L}_{0}$. Without any additional approximations, the resulting state is
\begin{equation} \label{eq:fstate}
\lim_{N \rightarrow \infty} \rho_f = |a|^2  r_{++}^0 +|b|^2 r_{--}^0 + \gamma_f a^* b r_{+-}^0 + \gamma_f ^*  a b^* r_{-+}^0, \qquad \gamma_f = \text{Tr}[ (\tilde{l}_{+-}^{0})^\dagger r_{+-}^0 ] \text{Tr}[ (l_{+-}^{0})^\dagger \tilde{r}_{+-}^0 ].
\end{equation}
We see that the final state is very simply related to the initial state via  the $\gamma_f$ parameter in Eq.~\eqref{eq:fstate}. Moreover, numerically we observe that $\gamma_f$ approaches $1$ exponentially fast in the thermodynamic limit, depicted in Fig.~\ref{fig:gamma-abs} for both the case of (a) $\kappa_d \neq 0$ and (b) $\omega \neq 0$. (We have also checked that $\gamma_m$ approaches 1 in the same limit.) This implies that the final state $\rho_f$ is indeed expected to return to its initial (pure) state $\rho_i$ in the thermodynamic limit.

\subsection*{Structure of the left eigenoperators $\tilde{l}$}

In Sec.~3 and earlier in this Section, we saw that the initial state settles into the noiseless subsystem of the intermediate Lindbladian  $\mathcal{L}_0+\mathcal{L}'$ without losing any coherences as $N\rightarrow \infty$. We would like to find a simple explanation for this behavior. This evolution would be accounted for (in the limit $N \rightarrow \infty$) if the left eigenoperators of $\mathcal{L}_{0} +\mathcal{L}'$ with zero eigenvalue are equal to the identity in each symmetry sector, since, in this case, $\gamma_m= \text{Tr}[ (l_{+-}^{0})^\dagger \tilde{r}_{+-}^0 ] = \text{Tr}[  \tilde{s}_{+-}^0 ]= \text{Tr}[  \tilde{z}_{+-}^0 ]=1$ where in the last step we have used $\text{Tr}[\tilde{z}_{+-}^0 ] = \text{Tr}[\tilde{z}_{++}^0 ] = 1$. (See Sec.~2 for definitions of $r,s,z$.)  We will show that this is indeed true. Splitting up the left eigenoperators into symmetry sectors, we have 
 \begin{equation} 
\tilde{l}_{++}^F = \left(\begin{array}{cc}
y_{++} & 0\\
0 & 0
\end{array}\right),\qquad
\tilde{l}_{--}^F = \left(\begin{array}{cc}
0 & 0\\
0 & y_{--}
\end{array}\right),\qquad
\tilde{l}_{+-}^F = \left(\begin{array}{cc}
0 & y_{+-}\\
0 & 0
\end{array}\right),\qquad
\tilde{l}_{-+}^F = \left(\begin{array}{cc}
0& 0\\
y_{-+} & 0
\end{array}\right).
\end{equation}
As before, we are in  the Fock basis  $[ \ket{0},\ket{2}, \ket{4},\ldots,  \ket{1},\ket{3}, \ket{5},\ldots ]^T$. Then  $y_{++} = y_{--} = \mathbb{I}$ since any arbitrary initial state must have unit overlap with the steady-state solutions with non-zero trace. Now we  switch from the Fock basis to the diagonal basis of $r$,  $r_i^d = U^\dagger r_i^F U$, $\tilde{l}_i^d = U^\dagger \tilde{l}_i^F U$, and obtain 
 \begin{equation}
\tilde{l}_{++}^d = \left(\begin{array}{cc}
q_{++} & 0\\
0 & 0
\end{array}\right),\qquad
\tilde{l}_{--}^d = \left(\begin{array}{cc}
0 & 0\\
0 & q_{--}
\end{array}\right),\qquad
\tilde{l}_{+-}^d = \left(\begin{array}{cc}
0 & q_{+-}\\
0 & 0
\end{array}\right),\qquad
\tilde{l}_{-+}^d = \left(\begin{array}{cc}
0& 0\\
q_{-+} & 0
\end{array}\right).
\end{equation}
Again, $q_{++} = q_{--} = \mathbb{I}$; we shall now probe the structure of the off-diagonal matrix $q_{+-}$.

In this basis, the four \textit{right} eigenoperators $r$  of $\mathcal{L}_{0} +\mathcal{L}'  $ with zero eigenvalue are just a single diagonal matrix $z$ in each of the four symmetry quadrants in the thermodynamic limit (see Sec.~2). This matrix $z$ is not pure, and in principle has infinite rank although  its eigenvalues fall off exponentially quickly as a function of the index, i.e.\  $z_{jj} \sim e^{-c j}$ for some $c>0$. In the case of a noiseless subsystem with full rank  $z$, Ref.~\cite{viola2010} proved that the corresponding conserved quantity must be the identity in each symmetry sector for a finite-dimensional Hilbert space. Since our bosonic model has an infinite-dimensional Hilbert space, these results do not immediately apply. Nevertheless, we numerically show that the conserved quantities approach the identity in the thermodynamic limit.

\begin{figure}[!]
    \centering
    \includegraphics[scale=0.3]{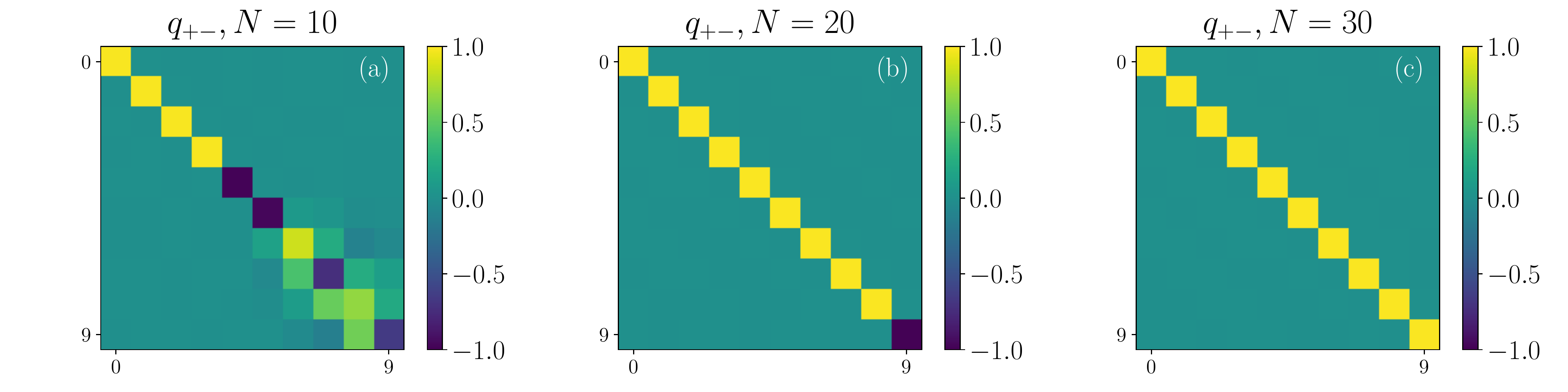}
    \caption{Plot of  a $10 \times 10$ block of  $q_{+-}$; all elements are real. Parameters: $\kappa_2 / \lambda = 1/N, \kappa_d / \lambda = 0.03, \omega = \kappa_1=0$. As the system approaches the thermodynamic limit, the matrix tends to the identity.  }
    \label{fig:qdfdep}
\end{figure}

\begin{figure}[!]
    \centering
    \includegraphics[scale=0.3]{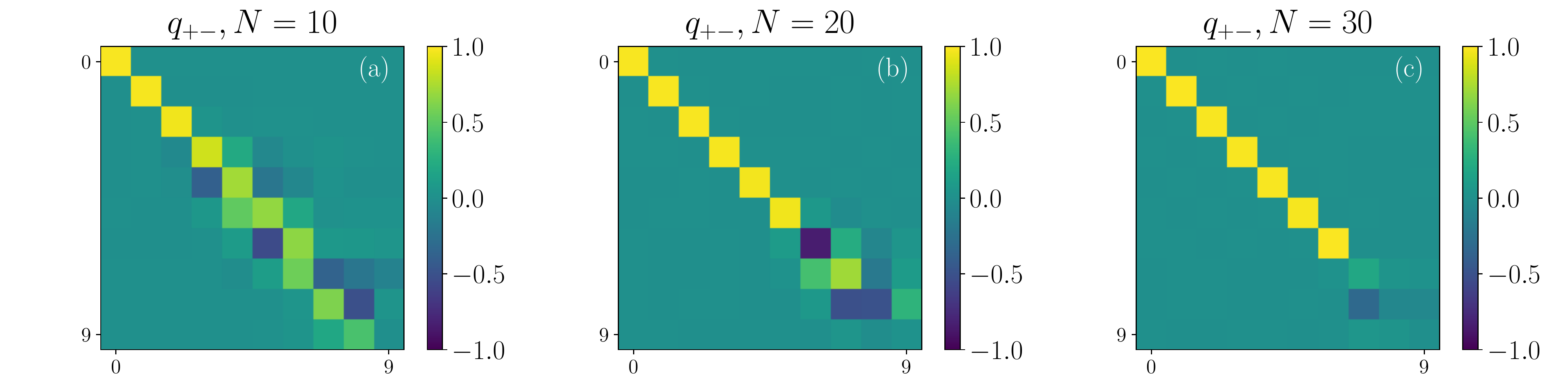}
  \caption{Plot of  a $10 \times 10$ block of  $q_{+-}$; all elements are real.  Parameters: $\kappa_2 / \lambda = 1/N, \omega/ \lambda = 0.5, \kappa_d = \kappa_1=0$. As the system approaches the thermodynamic limit, the matrix tends to the identity.  }
    \label{fig:qdfom}
\end{figure}

In Fig.~\ref{fig:qdfdep}, we plot the elements of  a $10 \times 10$ block of the matrix $q_{+-}$ for the case of non-zero dephasing. Indeed, we find that the matrix tends to the identity as we approach the thermodynamic limit. The matrix $q_{+-}$ acquires  off-diagonal terms at entries where  the corresponding matrix elements $z'_{jj}$ are small, i.e.~we are limited by numerical precision. Analogous behavior is observed for the case of non-zero $\omega$, depicted in Fig.~\ref{fig:qdfom}. So indeed we expect $\lim_{N \rightarrow \infty} q_{+-}  = \mathbb{I}$ for the full rank noiseless subsystem. This explains why $\rho_i$ does not lose coherences when relaxing into the steady state of $\mathcal{L}_{0} +\mathcal{L}'  $.

\end{document}